\documentclass[aps,prc ,twocolumn, nofootinbib,superscriptaddress]{revtex4-2}

\usepackage{slashed}
\usepackage{graphicx}
\usepackage{dcolumn}
\usepackage{bm}
\usepackage[dvipsnames,usenames]{xcolor}
\usepackage{amsmath}
\usepackage{amssymb} 
\usepackage{hyperref}
\usepackage{bbold}
\usepackage{braket}
\usepackage{stackengine}

\newcommand{\vecq}{{\bm q}}
\newcommand{\cL}{\mathcal{L}}
\newcommand{\cM}{\mathcal{M}}
\newcommand{\cA}{\mathcal{A}}
\newcommand{\cP}{\mathcal{P}}

\newcommand{\finalsinglet}{724.07 \pm 5.45}
\newcommand{\finaltriplet}{11.55 \pm 0.18}

\begin{document}	

\title{Muon Capture on the Proton with Heavy-Light Currents}

\author{Evan Combes}
\affiliation{Department of Physics and Astronomy, University of
  Tennessee, Knoxville, TN 37996, USA} 

\author{Emanuele Mereghetti}

\affiliation{Theoretical Division, Los Alamos National Laboratory, Los Alamos, NM 87545, USA}

\author{Lucas Platter}
\affiliation{Department of Physics and Astronomy, University of
  Tennessee, Knoxville, TN 37996, USA} \affiliation{Physics Division,
  Oak Ridge National Laboratory, Oak Ridge, TN 37831, USA}

\begin{abstract}

We construct a non-relativistic Lagrangian that describes muon-proton electroweak interactions. We determine the leading order coefficients by matching onto the theory with non-relativistic nucleons and relativistic leptons. The most impactful $\mathcal{O}(\alpha)$ corrections to those coefficients are determined by matching the non-relativistic amplitudes for capture of a free muon on a proton to the corresponding relativistic one. We use our non-relativistic effective field theory framework to calculate the capture rate in muonic hydrogen, thereby including radiative corrections of order $\alpha$ and up to order $1/m_\mu$. Using results for the Fermi coupling previously derived in EFT, we obtain singlet and triplet capture rates of $\Gamma({}^1 S_0) = \finalsinglet~\rm{s}^{-1}$ and $\Gamma({}^3 S_1) = \finaltriplet~\rm{s}^{-1}$, respectively. 
\end{abstract}

\maketitle

\section{Introduction}
\label{sec:intro}
Effective field theories (EFTs) have become a go-to tool in nuclear physics to describe observables in nuclear systems, such as bound state properties and reaction rates in electroweak processes and nuclear decays~\cite{Hammer:2019poc}. EFTs are systematic low-energy expansions with the expansion parameters being ratios of well separated scales that are inherent to the nuclear system. Two prominent EFTs in low energy nuclear theory are the pionless EFT, that describes the nuclear interaction using only contact interactions, and chiral EFT, that uses contact interactions and pion exchanges~\cite{Epelbaum:2008ga,Hammer:2019poc}. In an EFT approach, a number of parameters have to be calculated from the underlying full theory or obtained from experiment to acquire predictive power at a given order. Furthermore, experimental determinations frequently require that observables are sufficiently {\it different} to constrain the parameters of the Hamiltonian.

In recent years, significant progress has been made in describing electroweak reactions with high accuracy using EFTs. In using an EFT approach to that end, calculating relevant observables requires construction of the electroweak current. After that construction, a number of parameters that depend on the order of the calculation must then be obtained from experiment. Several observables now have reached the level of experimental and theoretical accuracy that demand the inclusion of radiative corrections for the extraction of Standard Model parameters and to search for physics beyond the Standard Model. Examples include the lifetime of superallowed $0^+ \rightarrow 0^+$ decays and electron energy spectra of pure Fermi and Gamow-Teller transitions. The description of these observables in the EFT framework usually starts in the single-nucleon sector \cite{Glick-Magid:2021uwb,King:2022zkz,Cirigliano:2024msg,Sargsyan:2026ygi},
and recently significant progress has been made in the description of neutron $\beta$-decay \cite{Ando:2004rk,Seng:2018qru,Seng:2018yzq,Hill:2023acw,Cirigliano:2022hob, Cirigliano:2023fnz,Cirigliano:2024nfi, VanderGriend:2025mdc,Cao:2025lrw,Tomalak:2026wks}. As single nucleon properties are a key component of currents used in many-body calculations, it is clear that the EFT description of the electroweak properties of few- and many-body systems is inherently connected to the description of single nucleon systems.

Another electroweak reaction type that has received a lot of attention is muon capture on nuclei, which has long served as fertile ground for the study of weak interactions and nuclear structure~\cite{Measday:2001yr}. In the EFT, it is another probe of the Hamiltonian and the electroweak current that can also be used to constrain low energy constants in a complementary fashion to $\beta$-decay, as the involved energy and momentum scales are frequently very different. Muon capture has been studied extensively in chiral perturbation theory \cite{Bernard:1995dp,Bernard:2000et,Ando:2000zw,Ando:2001uh}, where the individual contributions to the capture rate are inherently tied to a chiral description of the various single nucleon form factors. Furthermore, muon capture on the deuteron has received continuous interest as a process that can constrain the electroweak two-body current and thereby also the three-nucleon force. Additionally, muon capture on the deuteron has been considered as a way of determining the strength of the electroweak two-body current operator $L_{1A}$~\cite{Bonilla:2025uoo,Gnech:2023mvb}. Similarly, experimental measurement of muonic hydrogen weak capture reactions has in recent times been used for high precision determination of the proton's pseudoscalar coupling~\cite{MuCap:2007tkq} and thereby the axial radius. For such capture reactions it is therefore worthwhile to study the impact of radiative corrections that arise from the electromagnetic interaction between a muon and nucleus or nucleon. In this work we will focus on muonic hydrogen weak capture reactions.

Bound states, however, are notoriously difficult to work with in relativistic quantum field theories. This follows as a consequence of the presence of multiple energy scales and therefore motivates the use of a non-relativistic framework. This was already recognized by Caswell and Lepage~\cite{Caswell:1985ui}, who invented non-relativistic quantum electrodynamics (NRQED) in order to calculate Lamb shift and hyperfine structure effects in bound states of leptons. Likewise, we will here create a new non-relativistic framework based on NRQED using heavy-to-light currents which will allow for a comparatively painless description of electroweak interactions from hydrogenic atoms. Here, we apply this new framework to weak capture in muonic hydrogen and calculate capture rates for the spin-singlet and spin-triplet configurations. However, due to its structure, it will also be useful for muon capture reactions on nuclei where nuclear structure effects become relevant.


This manuscript is ordered as follows. We will begin by introducing the relevant Lagrangian from NRQED and then discuss the leading pieces of the Lagrangian representing heavy-to-light currents that will facilitate the use of a non-relativistic charged lepton in the relevant box diagrams. In Sec.~\ref{sec:matching}, we will match the low-energy coefficients appearing in this Lagrangian to a one-nucleon calculation that employs a non-relativistic description of the nucleon and a relativistic description of the muon. In Sec.~\ref{sec:magandpseudo} we will consider additional contributions to the matching coefficients arising from pion-photon loops.
We will then discuss the results obtained for muon capture in muonic hydrogen and conclude with a summary.

\section{Lagrangians}
\label{eq:Lagrangians}
The non-relativistic description of the interaction between charged particles and photons can be formulated using NRQED~\cite{Caswell:1985ui}. Its Lagrangian is well understood and has recently been constructed up to order $1/m^4$~\cite{Hill:2012rh}. Here, we limit ourselves to the terms up to order $1/m^2$, so that
\begin{multline}
\hspace{-0.2cm}\cL_{\rm NRQED} = \psi^{\dagger} \left[ iD_t + c_F \frac{q \bm{\sigma}\cdot \bm{B}}{2m} + c_D \frac{q (\bm{D} \cdot \bm{E} - \bm{E} \cdot \bm{D})}{8m^2} \right. \\
+ \left. c_S \frac{iq \bm{\sigma} \cdot (\bm{D} \times \bm{E} - \bm{E} \times \bm{D})}{8m^2} \right]  \psi~,
\end{multline}
where $m$ and $q$ are the particle mass and charge; $c_F$, $c_D$, and $c_S$ (``Fermi'', ``Darwin'', and ``spin-orbit'') are general coefficients which require matching; $\psi$ is the two-component spinor field of the particle; $\bm{D}$ is the space-like component of the covariant derivative; and $\bm{E}$ and $\bm{B}$ are the electric and magnetic fields. The low-energy coefficients can be obtained by matching to relativistic quantum electrodynamics (QED) and are expressed as expansions in powers of the fine structure constant $\alpha$. For point-like particles such as the muon, at leading order in the matching, one finds $c_F = c_D = c_S = 1$.

To describe muon capture, we start from a theory with relativistic muons interacting with non-relativistic nucleons, and then integrate out the scale of the muon mass to match onto NRQED.
As the nucleon mass is much larger than the muon mass, we can first integrate out the scale $m_N$
and describe the muon weak interactions with pions and nucleons 
using Heavy Baryon Chiral Perturbation Theory (HB$\chi$PT) \cite{Jenkins:1990jv, Bernard:1995dp}. At leading order, the Lagrangian is given by
\begin{align}\label{eq:HBweakLag}
    \cL_{\mu N}  &= -\sqrt{2}G_F V_{ud}  \bar{\Psi}_{\nu} \gamma^\mu P_L \Psi_\mu \, \left[
    \bar n (g^{}_V v_\mu - 2 g^{}_A S_\mu) p  \right. \nonumber\\
    &\hspace{0.5cm} \left. + \sqrt{2}  F^{}_\pi  \partial_\mu \pi^+ \right] + \textrm{h.c.}~,
\end{align}
where $\Psi_\nu$ and $\Psi_\mu$ are the relativistic neutrino and muon fields, $\pi^+$ is the pion field, and $p$ and $n$ are heavy baryon fields for the proton and neutron. $P_L = (1-\gamma_5)/2$ is the left-handed projector, while  $v^\mu = (1,0)$
and $S^{\mu} = (0, \vec{\sigma}/2)$. 
$V_{ud}$ is the up-down CKM matrix element, $V_{ud} = 0.9737(3)$ \cite{ParticleDataGroup:2024cfk}, currently extracted from superallowed $\beta$ decays, and $G_F$ is the Fermi constant, $G_F = 1.1663788\cdot 10^{-5}~\rm{GeV}^{-2}$~\cite{MuLan:2012sih}, determined from muon decay. 
$F_{\pi}$, $g_V$ and $g_A$ are the pion decay constant and the nucleon vector and axial couplings. 
Formally they should be interpreted to be given in the chiral limit and without electromagnetic corrections. Chiral and electromagnetic corrections are then included in a perturbative expansion, receiving contributions from both long-distance pion and photon loops and from low-energy constants in subleading HB$\chi$PT Lagrangians \cite{Knecht:1999ag,Gasser:2002am,Descotes-Genon:2005wrq,Ando:2004rk,Cirigliano:2023fnz}. Here we provide the physical values for these couplings, and make sure to avoid double counting when considering photon loops.
For the pion decay constants, we use $\sqrt{2}F_\pi = 130.41(20)~\rm{MeV}$~\cite{ParticleDataGroup:2024cfk}. 
At $\mathcal O(\alpha)$ the nucleon vector and axial couplings
$g_V$ and $g_A$ are both scale dependent, with a scale dependence that cancels the one arising from the photon loops showed in Fig.~\ref{fig:triangle-nr}. We find it convenient to express observables in terms of $g_V$, which has been studied very well from the theoretical point of view     \cite{Seng:2018qru,Czarnecki:2019mwq},
and of the ratio of the vector and axial couplings, $g_A/g_V$, which is scale independent and for which the electromagnetic corrections are subsumed in the experimental extraction from neutron decay,
$g_A/g_V = 1.2754 \pm 0.0013$~\cite{ParticleDataGroup:2024cfk}.  
Because of vector current conservation,
the vector coupling $g_V =1$ at leading order in $\alpha$. 
At $\mathcal O(\alpha)$, $g_V$ gets contributions from several energy scales, from the weak scale, $\mu_{\text{ew}} \sim M_W$, to the hadronic scale $\mu_{\text{had}} \sim m_N$\cite{Sirlin:1977sv}.
In our evaluation, we use updated results for the perturbative evolution between $\mu_{\text{ew}}$ and $\mu_{\text{had}}$ \cite{Cirigliano:2023fnz}, and the recent re-evaluations of the nonperturbative $W\gamma$ box \cite{Seng:2018qru,Seng:2018yzq,Czarnecki:2019mwq,Shiells:2020fqp,Hayen:2020cxh,Seng:2020wjq,Cirigliano:2022yyo}.
In the $\overline{\text{MS}}_\chi$ scheme~\cite{Gasser:1983yg}, which we will define later, and at the renormalization scales $\mu_\chi = m_N$ and $\mu_\chi= m_\mu$, $g_V$ is found to be \cite{Cirigliano:2023fnz} \begin{eqnarray}
    g_V(\mu_\chi = m_N) &= 1.01153(12) ~,\\
    g_V(\mu_\chi = m_\mu) &= 1.01545(12)~,
\end{eqnarray}
with error dominated by the nonperturbative $\gamma W$ box. We do not include
recent lattice QCD evaluations of the $\gamma W$ box
\cite{Ma:2023kfr} and
$\mathcal O(\alpha \alpha_s)$ corrections \cite{Moretti:2025ghw}, which lead to a small reduction in the error. The theoretical uncertainty on $g_V$ leads to a relative error on the muon capture rate of about $2.4 \cdot 10^{-4}$. This theoretical uncertainty could be avoided by normalizing the muon capture rate by the neutron decay rate, at the cost of introducing the experimental error on the latter, which is about $4 \cdot 10^{-4}$ \cite{ParticleDataGroup:2024cfk}, and thus of similar size.

At leading order in the electromagnetic coupling $\alpha$, one can use the Lagrangian in Eq.~\eqref{eq:HBweakLag}, and higher-order corrections, to express the capture amplitude in terms of a one-body hadronic current, $(J_{\mu}^-)^{1b}$, which contains a vector and axial component: $(J_{\mu}^-)^{1b} = V_{\mu}^- - A_{\mu}^-$.
The currents can be expressed in terms of vector electric and magnetic, axial and induced pseudoscalar form factors. The vector form factors are related by isospin symmetry to the isovector electromagnetic form factors, which can be extracted from data.  
Using phenomenological parameterizations that fit well to electron scattering data \cite{Alberico:2008sz}, one can check that, to achieve better than permille accuracy, it is sufficient to retain terms up to the isovector charge radius.  
While extractions of the axial form factor are more uncertain, fits to dipole parameterizations yield an axial mass of about 1 GeV, which justifies expanding also the axial form factor in $Q^2$ \cite{Bernard:2001rs}. The situation is different for the induced pseudoscalar form factor, where pion physics is important and we cannot expand in $m_\mu^2/m_\pi^2$. We thus express the currents as: 

\begin{align}
	V_0^{} &= N^\dagger \frac{\tau_a}{2}\left( 1 + \frac16 \braket{r_c^2} \overline{\nabla}^2 - \frac{\overline{\nabla}^2}{8m_N^2} \right) N ~,\label{eq:V0}\\
	A_0^{} &=  \frac{i}{2} N^\dagger g_A \tau_a \frac{\bm{\sigma} \cdot \overleftrightarrow{\nabla}}{2m_N} N \nonumber \label{eq:A0} \\
    &\hspace{0.5cm}- \frac{1}{4m_N^2}N^\dagger G_P (\overline{\nabla}^2) \overline{\partial}_0 \bm{\sigma} \cdot \overline{\nabla} \frac{\tau_a}{2}N ~,
\end{align}
while the spatial parts are
\begin{align}
	V_k^{} &=  \frac{i}{2} N^\dagger \frac{\overleftrightarrow{\nabla}_k }{m_N} \frac{\tau_a}{2} N \nonumber \\
    &\hspace{0.5cm} - N^\dagger  \kappa^{(1)} \tau_a  \epsilon_{kij} \frac{\sigma_i \overline{\nabla}_j }{2m_N} N \left( 1 + \frac16 \langle r^2_M \rangle \overline{\nabla}^2 \right) ~,
    \label{eq:Vk} \\
	A_k^{} &=  \frac{1}{2} N^\dagger g_A \tau_a \sigma^k \left( 1 + \frac{1}{6} \braket{r^2_A}\overline{\nabla}^2 - \frac{\overline{\nabla}^2}{8m_N^2} \right) N \nonumber\\
    &\hspace{0.5cm}+  \frac{1}{4m_N^2} N^\dagger G_P (\overline{\nabla}^2) \overline{\nabla}_k  \bm{\sigma} \cdot \overline{\nabla}  \frac{\tau_a}{2}N~\label{eq:Ak}~.
\end{align}
Here we give for completeness also $A_0$ and $V_k$. As these contributions scale as $\mathcal{O}(\frac{1}{m_N})$, we will not consider radiative corrections to their contribution.
Above, $N$ denotes the non-relativistic nucleon field, the operator $\overline{\nabla}$ is defined as $\overline{\nabla} = (\overleftarrow{\nabla} + \overrightarrow{\nabla})$ and likewise, $\overline{\partial}_0 = (\overleftarrow{\partial}_0 + \overrightarrow{\partial}_0)$; conversely, $\overleftrightarrow{\nabla} = (\overleftarrow{\nabla} - \overrightarrow{\nabla})$. $\kappa^{(1)}$ is the nucleon isovector magnetic moment, $\kappa^{(1)} = \frac12(\kappa_p - \kappa_n) = 2.35295$.
$\langle r^2_c \rangle$, $\langle r^2_A \rangle$, and $\langle r^2_M \rangle$
are the isovector nucleon charge,  axial, and magnetic radii, and $G_P$ is the induced pseudoscalar form factor.
These quantities have an expansion in HB$\chi$PT 
(see, for example,
Ref. \cite{Bernard:1995dp}
and references within),
or can be expressed in terms of experimentally accessible nucleon form factors. In this paper we will use the world average experimentally measured values for the proton and neutron charge radii, $\sqrt{\langle r^2_c \rangle_p} = 0.84075 \pm 0.00064~\text{fm}$ and $\langle r^2_c \rangle_n = -0.1155 \pm 0.0017~\text{fm}^2$, respectively, to produce the isovector charge radius $\langle r^2_c \rangle = 0.8224 \pm 0.0020~\text{fm}^2$. For the axial radius we use $\langle r^2_A \rangle = 0.369 \pm 0.030~\text{fm}^2$, obtained from LQCD~\cite{MINERvA:2025ygc,Meyer:2026kdl}. Lastly, for the magnetic radius we use the world average measurements for the proton and neutron magnetic moments and magnetic radii to determine $\langle r^2_M \rangle = 0.733 \pm 0.027~\rm{fm}^2$.
For power counting purposes, we recall that in HB$\chi$PT
both $\langle r^2_c \rangle$ and $\langle r^2_A \rangle$ scale as $\Lambda_\chi^{-2}$,
where $\Lambda_\chi \approx 4 \pi F_\pi \approx m_N$ is the HB$\chi$PT breakdown scale.\\

The pseudoscalar form factor is, at lowest order,
\begin{equation}\label{eq:GP}
    G_P^{(0)}(q^2) = \frac{F_\pi g_{np \pi^\pm}}{m_N} \frac{4m_N^2}{ m_{\pi^\pm}^2 - q^2} ~,
\end{equation}
where $g_{np \pi^\pm}$ is the pion-nucleon coupling. With the Goldberger-Treiman relation, and the dominance of the pion pole, $G_P^{(0)}(q^2)$ can be approximated at leading order as
\begin{equation}
    G_P^{(0)}(q^2) \approx g_A \frac{4m_N^2}{ m_{\pi^\pm}^2 + \bm{q}^2 }~.
\end{equation}
In the numerical analysis we use Eq.~\eqref{eq:GP}, with $g_{np \pi^\pm} = 13.12(10)$~\cite{Hill:2017wgb}. This value is compatible with the more recent extraction of Ref.~\cite{Reinert:2020mcu}, which yields $g_{np \pi^\pm} = 13.23(4)$. 
At higher order in $\alpha$, the muon-nucleon interactions cannot be purely described in terms of single nucleon form factors. At $\mathcal O(\alpha)$
and $\mathcal O(\alpha m_\mu/\Lambda_\chi)$, 
the effects of two and three weak and electromagnetic currents acting on hadronic states need to be considered, leading to both long-range photon-exchange diagrams, as the ones shown in Figs. \ref{fig:triangle-nr} 
and \ref{fig:pseudo}, and new contact interactions, which parameterize the short-distance behavior of similar diagrams \cite{Gasser:2002am,Descotes-Genon:2005wrq,Ando:2004rk,Cirigliano:2022hob,Cirigliano:2023fnz,Cirigliano:2024nfi}. 
The counterterm Lagrangians needed for the calculation at $\mathcal O(\alpha)$ and
the representation of the low-energy constants in terms of nonperturbative QCD correlation functions were derived in  
Refs. \cite{Gasser:2002am,Descotes-Genon:2005wrq,Ando:2004rk,Cirigliano:2022hob,Cirigliano:2023fnz,Cirigliano:2024nfi}. The $\mathcal O(\alpha m_\mu/\Lambda_\chi)$
Lagrangian has not been constructed yet. This would be important to increase the accuracy of the EFT calculation.

With the nucleon mass scale integrated out, we next integrate out the scale of the muon mass.
In addition to  
non-relativistic muons and baryons and 
soft and ultrasoft photons and neutrinos,
the resulting effective theory contains highly energetic ``collinear'' photons and neutrinos, with 
one component of the momentum of order of the muon mass.
The construction of the transition operator is thus analogous to that of heavy-light currents in Heavy Quark Effective and Soft Collinear Effective Theory
\cite{Bauer:2000ew,Bauer:2000yr,Fontes:2025xbt}, with the simplification that, owing to the neutrality of neutrinos, 
collinear photons do not appear at leading order in $1/m_\mu$ \cite{Bauer:2000ew,Bauer:2000yr}.
At leading order in $1/m_\mu$, the non-relativistic Lagrangian has the form
\begin{align}\label{eq:nreftlagrangian1}
    \cL_{\mu N}^{\rm NR(0)} = -\frac{1}{\sqrt{2}} G_F V_{ud} g_V \bar{\Psi}_\nu \Omega^\beta  \psi_\mu  \left( \mathcal V_\beta -\mathcal A_\beta\right) + \textrm{h.c.}~,
\end{align}
where the symbols $\mathcal V_{\beta}$ and $\mathcal A_{\beta}$ denote that these operators originate, at tree level, from the vector and axial currents, even if this distinction is lost at higher orders in $\alpha$.
As before, $\Psi_\nu$ is the relativistic neutrino field, while now $\psi_\mu$ is the non-relativistic muon field. The operator $\Omega^\beta$ is defined as $\Omega^\beta = \begin{pmatrix}
    \Omega^0,~-\Omega^j
\end{pmatrix}$, where 
\begin{equation}
    \Omega^0 = \begin{pmatrix} \sigma^0 \\ \sigma^0 \end{pmatrix}~,~ \Omega^j = \begin{pmatrix} \sigma^j \\ \sigma^j \end{pmatrix}~.
\end{equation}
The $0$th components of the hadronic operators in Eq.~\eqref{eq:nreftlagrangian1} 
are given by
\begin{align}
\mathcal V_0 &= w_1(m_\mu)  n^\dagger p ~, \label{eq:V0low}     \\
\mathcal A_0 &=  w_2(m_\mu) i  \, \frac{g_A}{g_V}  n^\dagger \frac{\bm{\sigma} \cdot \overleftrightarrow{\cP}}{2m_N} p \nonumber \\
    &\hspace{0.5cm} + w_3(m_\mu) n^\dagger \overline{\mathcal P}_0 \bm{\sigma} \cdot \overline{\mathcal{P}} p ~, 
\end{align}
while the space-like components are given by
\begin{align}
\mathcal V_k &= w_4(m_\mu) 
n^\dagger  \frac{i \overleftrightarrow{\cP}_k }{2m_N} p  - w_5(m_\mu) n^\dagger \epsilon_{kij} \frac{\sigma_i \overline{\cP}_j }{2m_N} p ~,
\\
\mathcal A_k &= \frac{g_A}{g_V} w_6(m_\mu)
 n^\dagger \sigma^k p +  w_7(m_\mu) n^\dagger  \overline{\cP}_k  \bm{\sigma} \cdot \overline{\cP}  p ~, \label{eq:Aklow}
\end{align}
where the operators $\overline{\cP}$, $\overleftrightarrow{\cP}$ are defined the same as for Eqs.~\eqref{eq:V0}--\eqref{eq:Ak}, with the difference that here the derivatives select only the large component of the momentum that scales as $m_\mu$. Likewise, $\overline{\cP}_0$ selects only the large component of the energy that scales as $m_\mu^2/m_N$. Thus, the matching coefficients $w_i$ are determined at the muon mass scale and hence are explicitly written as $w_i(m_\mu)$. In practice, contributions from the small component of momentum could be calculated explicitly, though for consistency in the order of our calculations, they are not included in this work.
At tree level, the matching coefficients are 
\begin{align}
w_1 &= 1 - \frac{1}{6} \langle r^2_c \rangle m_\mu^2 + \frac{m_\mu^2}{8m_N^2} \label{eq:M1} ~,\\
w_2 &= 1 ~,\\
w_3 &= \frac{g_{np \pi^\pm} F_\pi}{m_N} \frac{1}{m_\pi^2 + m_\mu^2}~, \\
w_4 &= 1 ~,\\
w_5 & = 2 \kappa^{(1)}\left(1 - \frac16 \langle r^2_M \rangle m_\mu^2  \right)  \label{eq:M2} ~,\\
w_6  &= 1 - \frac{1}{6} \langle r^2_A \rangle m_\mu^2 + \frac{m_\mu^2}{8m_N^2} \label{eq:M3}~, \\
w_7 &= \frac{g_{np \pi^\pm} F_\pi}{m_N} \frac{1}{m_\pi^2 + m_\mu^2} - \frac{1}{6} \frac{g_A}{g_V} \langle r_A^2 \rangle ~, \label{eq:M4}
\end{align}
where we truncated the expressions at $\mathcal O(m^2_\mu/\Lambda_\chi^2)$.
We thus see that the matching coefficients have a polynomial expansion in 
$m_{\mu}/\Lambda_\chi$, 
and can have 
 an arbitrary dependence on all scales of
 order of the muon mass, e.g. $m_\pi$.
When we add electromagnetic effects, the  
expansion can be generalized as
\begin{align}\label{eq:ws}
    w_i = \sum_{n,m} w_i^{(n,m)}(m_\mu) \left(\frac{m_\mu}{\Lambda_\chi}\right)^n \alpha^m~.   
\end{align}
The goal of this paper is to compute the $\mathcal O(\alpha)$ corrections to the leading terms in the $m^{-1}_N$, 
$\Lambda^{-1}_\chi$
expansion of the single-nucleon currents. Therefore, the coefficients $w_1$ and $w_6$ will receive  $\mathcal{O}(\alpha)$ corrections, and corrections to $w_7$, where the factor of $m_N^{-1}$ is conventional,  are also calculated. Corrections to $w_2$, $w_4$, and $w_5$ contribute at order $\alpha {p}/{m_N}$ and are not considered here.

The lowest order Lagrangian \eqref{eq:nreftlagrangian1} can then be extended to higher order in $1/m_\mu$. 
At NLO, we will focus on the momentum-independent vector and axial operators, which are not suppressed by powers of the nucleon mass. We will write
\begin{align}\label{eq:nreftlagrangian2}
\nonumber
\mathcal L^{\text{NR} (1)}_{\mu N} &=     
-\frac{1}{\sqrt{2}} G_F V_{ud} g_V \bar{\Psi}_\nu \Omega^\beta \left[ i\frac{\bm{\sigma} \cdot \overrightarrow{\nabla}}{2m_\mu} \right] \psi_\mu \\
& \times\left[  \delta_{\beta 0} w_1^\prime(m_\mu) n^\dagger p
+
\delta_{\beta k} w_6^\prime(m_\mu) n^\dagger \sigma^k p \right]~.
\end{align}
However, due to the momentum structure in the leptonic current, the above terms do not contribute to the weak capture rates of singlet and triplet channel muonic hydrogen. Fortunately, $\mathcal{O}(\alpha)$ corrections to the coefficients $w_1^\prime$ and $w_6^\prime$ manifest in the process of determining $\mathcal{O}(\alpha)$ corrections to $w_1$ and $w_6$, and so they will still be presented here.

\section{Matching calculation} \label{sec:matching}
\begin{figure*}[t]
    \includegraphics[scale=0.85]{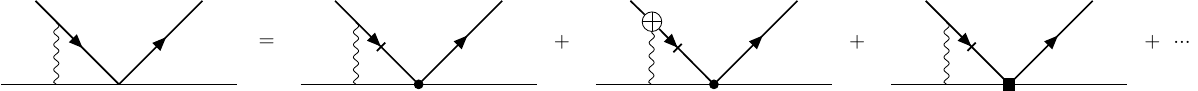}
\caption{Comparison between the relativistic triangle diagram and corresponding non-relativistic triangle diagrams for muon capture up to $\mathcal{O}\left( \alpha\frac{p}{m_\mu} \right)$. Lines with (without) arrows are leptons (nucleons). A dash with an arrow indicates a non-relativistic lepton. Wavy lines are Coulomb photons. 
On the weak vertex, black circles (squares) represent LO (NLO) weak leptonic interactions. Circles with a plus sign inside are NRQED Darwin vertices. Lepton-photon vertices without special indicator are NRQED Coulomb vertices.}
\label{fig:triangle-nr}
\end{figure*}
As stated, we carry out the matching of the $w_i$ coefficients by calculating the amplitude for muon capture using a relativistic description of the muon up to order $\alpha$. We then perform, in addition, a non-relativistic reduction by expanding results in powers of $|\bm{p}_\mu|/m_\mu$. Comparison to the corresponding calculations performed using the non-relativistic framework thus allows for determination of $\mathcal{O}(\alpha)$ corrections to the $w_i$ coefficients. In this part of the manuscript, we will focus explicitly on the Gamow-Teller contribution to the capture rate and thus identify $\mathcal{O}(\alpha)$ corrections to $w_6$ and $w_6^\prime$. The relevant triangle diagrams are displayed in Fig.~\ref{fig:triangle-nr}. As the momentum structure of the Fermi and Gamow-Teller interactions are the same, we anticipate the $\mathcal{O}(\alpha)$ corrections to $w_1$ and $w_1^\prime$ are the same as those to $w_6$ and $w_6^\prime$. Throughout the following discussion, we will use $\cM$ to indicate amplitudes calculated with relativistic muons and $\mathcal{M}^{\rm NR}$ for subsequent non-relativistic reductions in $1/m_\mu$ to those amplitudes. $\cA$ will be used for amplitudes calculated with non-relativistic muons.

It is instructive to first look at the tree level diagrams. The tree level diagram with Gamow-Teller weak vertex is given by
\begin{equation}
    \cM_{\rm tree} = \bar{u}_\nu \gamma^k P_L u_\mu (\mathcal{J}^-_k)^{(1b)}~,
\end{equation}
where $(\mathcal{J}^-_k)^{(1b)}$ is the hadronic component of the amplitude. Performing the above-mentioned non-relativistic reduction, one finds
\begin{align}
    \cM_{\rm tree}^{\rm NR} &= \bar{u}_\nu (-\Omega^k)  \left[ 1 - \frac{\bm{\sigma} \cdot \bm{p}_\mu}{2m_\mu}  \right]\chi^\mu (\mathcal{J}^-_k)^{(1b)}~,
\end{align}
where $\chi^\mu$ is the two-component spinor of the muon ($\mu$ is here not an index). 
We note that 
through the above non-relativistic reduction, we decomposed the tree level leptonic amplitude into a leading leptonic current and a subleading one that is of order $1/m_\mu$, in alignment with Eqs.~\eqref{eq:nreftlagrangian1},~\eqref{eq:nreftlagrangian2}. 

Having seen the tree level amplitudes, we now turn our attention to the triangle diagram amplitude with relativistic muon, seen in Fig.~\ref{fig:triangle-nr}. In handling electromagnetic interactions, we will operate in the Coulomb gauge, where correspondingly, the leading triangle diagram amplitudes entail exchange of Coulomb photons. Transverse photon exchanges do not contribute until $\mathcal{O}\left( \alpha \frac{p}{m_N}\right)$, and so are not covered in this work. The amplitude for the triangle diagram with relativistic muon is
\begin{multline}\label{eq:triangleamplitude}
    \hspace{-0.3cm}\cM_{\rm triangle} = (-e^2) (\mu^2)^{\frac{3-d}{2}} \int \frac{d^{d+1}q}{(2\pi)^{d+1}} \frac{1}{\vecq^2} \frac{i}{-q^0 - \frac{\vecq^2}{2m_p} + i\varepsilon} \\
    \times  \bar{u}_\nu \gamma^j P_L \frac{(\slashed{p}_\mu + \slashed{q} + m_\mu)}{(p_\mu + q)^2 - m_\mu^2 + i\varepsilon} \gamma^0 u_\mu~,
\end{multline}
where $e$ is the elementary charge, $\mu$ is the renormalization scale, and $d$ is the dimensional regularization dimension, $d \rightarrow 3$. Since the nucleon mass is larger than any other scale in the problem, we neglect the recoil energy $\vecq^2/2m_p$ of the proton. After solving and taking the non-relativistic reduction by expanding around zero muon momentum, we are left with
\begin{widetext}
\begin{align}\label{eq:triangle-rel-reduction}
	\cM_{\rm triangle}^{\rm NR} &= - (-e^2) \bar{u}_\nu \Omega^j \chi^\mu \left\{ \frac{1}{8\pi\beta_{\rm NR}}\left[  \frac{-2i}{3-d} + \pi + i\ln\left( \frac{\bm{p}_\mu^2}{4\pi \mu^2} \right) - i\psi^{(0)}\left( \frac12 \right) \right] + \frac{i\beta_{\rm NR}}{16\pi}  \right. \nonumber \\
	&\hspace{1cm} -  \left. \frac{1}{8\pi^2}\left[ \frac{-2}{3-d} + \gamma_E -\ln(4\pi) + \ln\left( \frac{m_\mu^2}{\mu^2}\right) + 2 \right] \right\} \nonumber \\
	&\hspace{0.5cm} + (-e^2) \bar{u}_\nu \Omega^j \frac{\bm{\sigma} \cdot \bm{p}_\mu}{2m_\mu}  \chi^\mu \left\{ \frac{1}{8\pi\beta_{\rm NR}}\left[  \frac{-2i}{3-d} + \pi + i\ln\left( \frac{\bm{p}_\mu^2}{4\pi \mu^2} \right) - i\psi^{(0)}\left( \frac12 \right) - 2i \right]  \right. \nonumber \\
	&\hspace{1cm} - \left. \frac{1}{8\pi^2}\left[ \frac{-2}{3-d} + \gamma_E - \ln(4\pi) + \ln\left( \frac{m_\mu^2}{\mu^2}\right) + 2 \right] + \frac{1}{6\pi^2} \right\}~.
\end{align}
\end{widetext}
Above, $\psi^{(0)}$ is the order-zero polygamma function and $\beta_{\rm NR} = |\bm{p}_\mu|/m_\mu$.

In addition to the relativistic triangle diagram contributions, another contribution that has to be accounted for at this order in $\alpha$ is the first order correction to the muon wave function renormalization. The Z-factor $Z_2$ at this order was calculated by Adkins in Ref.~\cite{Adkins:1982zk} in the Coulomb gauge. In their calculation, a logarithmic dependence of the  renormalization scale $\mu$ was eliminated by setting $\mu = m_\mu$. To recover this logarithm, we have, therefore, performed the same calculation with arbitrary renormalization scale and find
\begin{equation}
\label{eq:Z2}
    Z_2 = 1 - \frac{\alpha}{4\pi}\left[ \frac{2}{3-d} - \gamma_E + \log(4\pi) + \log\left(  \frac{\mu^2}{m_\mu^2} \right) \ \right]~.
\end{equation}
The complete expression up to order $\alpha$ for the capture amplitude is therefore
\begin{align}
    \cM^{\rm NR} = \sqrt{Z_2} \cM_{\rm tree}^{\rm NR} +\cM_{\rm triangle}^{\rm NR}~.
\end{align}
We need to reproduce this expression with diagrams constructed using the NRQED Lagrangian. The structure of Eq.~\eqref{eq:triangle-rel-reduction} hints that there are three requisite diagrams. The leptonic structures of Eq.~\eqref{eq:triangle-rel-reduction} indicate two diagrams with simple Coulomb vertices for both photon-muon and photon-proton interactions, but with leading and subleading leptonic currents in the weak interactions. For the third requisite diagram, the term linear in $\beta_{\rm NR}$ indicates a diagram with Darwin vertex for the photon-muon interaction, which is paired with the leading leptonic current. The three diagrams are shown in Fig.~\ref{fig:triangle-nr}. 

The three corresponding amplitudes are readily evaluated. We find the sum of the amplitudes to be 
\begin{widetext}
\begin{align}\label{eq:nrtriangle}
   \cA_{\rm triangle} &= -(-e^2) \bar{u}_\nu \Omega^j \chi^\mu\left[  w_6 \frac{1}{8\pi \beta_{NR}}\left\{ \frac{-2i}{3-d} + \pi + i\ln\left( \frac{\bm{p}_\mu^2}{4\pi \mu^2} \right) - i\psi^{(0)}\left( \frac12 \right) \right\}   + \frac{i \beta_{NR}}{16\pi} w_6 \right] \nonumber \\
	&\hspace{0.5cm} + (-e^2) \bar{u}_\nu \Omega^j \frac{\bm{\sigma} \cdot \bm{p}_\mu}{2m_\mu} \chi^\mu w_6^\prime \frac{1}{8\pi \beta_{NR}}\left\{ \frac{-2i}{3-d} + \pi + i\ln\left( \frac{\bm{p}_\mu^2}{4\pi \mu^2} \right) - i\psi^{(0)}\left( \frac12 \right) - 2i \right\} ~.
\end{align}
\end{widetext}
By matching $\cA_{\rm triangle}$ against the non-relativistic reduction to $\cM$, we determine coefficients:
\begin{align}
    w_6 = w_1 &= 1 + \frac{\alpha}{4\pi}\frac32 \ln\left( \frac{\mu^2}{m_\mu^2}\right) - \frac{\alpha}{\pi}~, \\
    w_6^\prime = w_1^\prime &= 1 + \frac{\alpha}{4\pi}\frac32 \ln\left( \frac{\mu^2}{m_\mu^2}\right) - \frac{\alpha}{3\pi} ~.
\end{align}
The scale dependence in the above equations is absorbed by a corresponding scale dependence in the vector coupling $g_V$. This scale dependence was determined in Ref. \cite{Ando:2004rk,Cirigliano:2022hob}. Note that though Eq.~\eqref{eq:triangleamplitude} is expressed with the minimal subtraction scheme in mind, the scheme-dependent terms $\log(4\pi)$ and $\gamma_E$ were subtracted as they would be in the $\overline{\text{MS}}$ scheme. As well, $g_V$ was determined using the $\overline{\text{MS}}_\chi$ scheme; for consistency, we let $\mu^2 = \mu^2_\chi e^{-1}$, where we use $\mu_\chi = m_\mu$.

\section{Corrections to the induced pseudoscalar form factor}\label{sec:magandpseudo}

\begin{figure}[b]
     \begin{minipage}[b]{0.1\textwidth}
        \includegraphics[width=\textwidth]{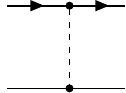}
    \end{minipage}
    \hspace{0.5cm}
    \begin{minipage}[b]{0.1\textwidth}
        \centering
        \includegraphics[width=\textwidth]{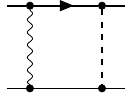}
    \end{minipage}
    \vspace{0.5cm}

    \hspace{-0.1cm}\begin{minipage}[b]{0.1\textwidth}
        \includegraphics[width=\textwidth]{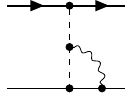}
    \end{minipage}
    \hspace{0.5cm}
    \begin{minipage}[b]{0.1\textwidth}
        \centering
        \includegraphics[width=\textwidth]{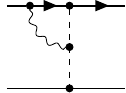}
    \end{minipage}

    \caption{Pion-photon loop diagrams representing the set of topologies which contribute at $\mathcal{O}(\alpha)$. Simple straight lines are nucleons, lines with arrows are leptons, dashed lines are pions, and wavy lines are photons.}
    \label{fig:pseudo}
\end{figure}

Another important contribution to the muon-proton weak capture rate arises from the induced pseudoscalar form factor of the nucleon,
which particularly impacts the spin-triplet capture rate. Indeed,
previous theoretical calculations show that while the weak-magnetic and pseudoscalar interactions have a minor impact on the singlet capture rate, they are responsible for a significantly larger triplet state capture rate than would otherwise be expected~\cite{PhysRevLett.99.032003}.

At leading order in HB$\chi$PT, the induced pseudoscalar form factor arises from a pion exchange between leptons and nucleons. 
Representative topologies of the pion-photon loop diagrams contributing at $\mathcal O(\alpha)$ are shown in Fig.~\ref{fig:pseudo}. The diagrams are ultraviolet and infrared divergent, and they are complicated by the appearance of three scales: $q^2$, $m_\mu^2$ and $m_\pi^2$. At leading order in the NRQED expansion, we can simplify the problem by taking $q^2 =  - m_\mu^2$. The ultraviolet divergences are absorbed by the electromagnetic renormalization of the charged pion mass, $m_{\pi^+}$ \cite{Gasser:2002am}, of the pion decay constant $F_{\pi^\pm}$ \cite{Knecht:1999ag,Descotes-Genon:2005wrq}, and of the proton-neutron-$\pi^+$ pion-nucleon coupling, $g_{np \pi^\pm}$ \cite{vanKolck:1997fu,Reinert:2022jpu,Reinert:2020mcu}.
All low-energy constants from the $\mathcal O(e^2 p^2)$ mesonic Lagrangian \cite{Gasser:2002am,Knecht:1999ag,Descotes-Genon:2005wrq}
and the $\mathcal O(e^2 p)$ nucleon-pion Lagrangian \cite{Gasser:2002am,Cirigliano:2022hob,Cirigliano:2023fnz} can be expressed in terms of these physical parameters. 
Some care has to be taken in order to define $F_{\pi^\pm}$ and 
the pion-nucleon coupling at $\mathcal O(\alpha)$.
For $F_{\pi^\pm}$, we use the definition of  Ref. \cite{Knecht:1999ag} for $F_\pi$ in the isospin limit. 
We then define $F_{\pi^\pm}$ to  include the contribution of $\mathcal O(e^2 p^2)$ low-energy constants, which are collectively denoted by $E^r$ in Ref. \cite{Knecht:1999ag}, and of the appropriate logarithm needed to make the coupling scale independent 
\begin{equation}
    F_{\pi^\pm} = F_\pi \left[ 1  + \frac{\alpha}{4\pi} \frac{1}{2} \left( (4\pi)^2 E^r(\mu_\chi) - 3 \log\frac{\mu_\chi^2}{m_\mu^2}  \right)\right]~. 
\end{equation}
This combination can be extracted from pion decay.
For the pion-nucleon coupling, we follow the prescription outlined in Refs.~\cite{vanKolck:1997fu,Reinert:2022jpu,Reinert:2020mcu}.
These papers calculated the pion-photon corrections to the nucleon-nucleon potential, which is given by diagrams very similar to those shown in Fig. \ref{fig:pseudo}, with the leptonic line replaced by another nucleon line. 
The loop would yield a correction to the isospin conserving nucleon-nucleon potential given by
\begin{align}
    V  &= - \frac{\alpha}{4\pi}  \left[\frac{g^{}_A}{2 F_\pi} \right]^2 \bm{\sigma}^{(1)} \cdot \vecq\, \bm{\sigma}^{(2)} \cdot \vecq\, \tau^{(1)} \cdot \tau^{(2)} \nonumber  \\ &  \times\Bigg\{
    \frac{1}{{\vecq}^{\,2} + m_\pi^2}
    \left[
         \kappa^r   + 2 \log \frac{\mu_\chi^2}{m_\pi^2} - 2 \right]
    + \frac{1}{m_\pi^2} V_{\pi \gamma}
    \Bigg\}~,
\end{align}
where $\kappa^r$ is a combination of the renormalized LECs $g_{1,2}$ and $\kappa_{1,2}$, defined in Ref. \cite{Gasser:2002am}, and
\begin{equation}\label{eq:Vpigamma}
    V_{\pi \gamma} = 4 \left( \frac{1}{2 \beta^2} - \frac{(1-\beta^2)^2}{2 \beta^4 (1+\beta^2)} \log (1+\beta^2)\right)~,
\end{equation}
where here, $\beta = |\vecq\, |/m_\pi$.
In these expressions, we have removed the corrections to the charged pion mass, which can be put into the tree level one-pion-exchange (OPE) potential.
Due to the logarithm in Eq.~\eqref{eq:Vpigamma}, it is not possible to unambiguously define the residue at the pion pole $|\vecq\,|^2 = - m_\pi^2$ ($\beta^2=-1$), and a scheme choice needs to be made to define the OPE. 
In Ref. \cite{Reinert:2020mcu}, the charged pion coupling is defined with the choice
\begin{equation}\label{eq:OPEscheme}
    \kappa^r + 2 \log \frac{\mu_\chi^2}{m_\pi^2} - 2 = - 8 \gamma_E~,
\end{equation}
which we also adopt here. 
This choice defines $g_{np \pi^\pm}$ at $\mathcal O(\alpha)$ and determines the combination of LECs $\kappa^r$ to be used in the induced pseudoscalar form factor at $\mathcal O(\alpha)$.

The difference between the relativistic and NRQED calculations yields corrections to the matching coefficients in Eqs. \eqref{eq:V0}--\eqref{eq:Ak}.
Again we find that the infrared divergences cancel between the relativistic and NRQED calculations. 
In addition, we find a correction to the axial couplings $w_6$  and to the pseudoscalar form factor $w_7$.
In the notation of Eq.~\eqref{eq:ws}, the correction to the axial form factor is
\begin{align}
   w^{(0,1)}_6(m_\mu) &= \frac{1}{4\pi} \Bigg\{ \frac{1}{2} -\frac{1}{2} \frac{1}{1-x_\mu} \log x_\mu \nonumber \\ &  -   h_{2}(x_\mu) + \frac{1 + x_\mu}{2 x_\mu}  f_{1}(x_\mu) \Bigg\}~,
\end{align}
where $x_\mu = m_\mu^2/m_\pi^2$ and
the loop functions are defined in Appendix~\ref{appendix:loopfunctions}.
In the limit of $x_\mu \ll 1$ this correction would scale as $m_\mu/m_\pi$ and thus is negligible for $\beta$ decays \cite{Cirigliano:2022hob}.
 As far as the pseudoscalar form factor is concerned, we find that radiative corrections shift $w_7$ as 
\begin{equation}
    w_7 = \frac{F_{\pi^+} g_{np \pi^\pm}}{m_N} \frac{1}{m^2_{\pi^\pm} + m_\mu^2} - \frac{1}{6} \frac{g_A}{g_V} \langle r_A^2 \rangle + \alpha  w^{(0,1)}_7~,
\end{equation}
with 
\begin{align}
 &   w^{(0,1)}_7 = -\frac{g_A}{4\pi m_\pi^2} \Bigg\{ 
\frac{4 (1+ \gamma_E)}{1+ x_\mu}
- \frac{1}{2 x_\mu (1+ x_\mu)} \nonumber \\ &     
 +   \frac{1}{2 x_\mu (1-x_\mu)} \log x_\mu  + \frac{x_\mu -3}{x_\mu (1+x_\mu)} \log (1+x_\mu)
 \nonumber \\ & - \frac{1+x_\mu}{2 x_\mu^2} f_1(x_\mu) 
+ \frac{4}{1+x_\mu} h_1(x_\mu) + \frac{1}{x_\mu} h_2(x_\mu) \nonumber \\ & - \frac{1-x_\mu+ 2x_\mu^2}{x^2_\mu (1+x_\mu)} h_3(x_\mu)  \Bigg\}~,
\end{align}
where the factor of $\gamma_E$ in the
first line is a consequence of the choice in Eq.~\eqref{eq:OPEscheme}.

Numerically, the shift to the axial form factor is 
\begin{equation}
   \alpha w^{(0,1)}_6 = -1.3 \cdot 10^{-3}~,
\end{equation}
close to the experimental error on $g_A$.
We can compare the correction to $w_7$ with the one arising from the axial radius. In our conventions, 
\begin{equation}
    w_7^{(2,0)} = - \frac{1}{6} \frac{g_A}{g_V} \langle r^2_A \rangle = - (2.02 \pm 0.16) \, \text{GeV}^{-2}~,
\end{equation}
using the Lattice QCD average of the axial radius discussed in Sec.~\ref{eq:Lagrangians}.
The contribution from the radiative corrections in Fig. \ref{fig:pseudo} is
\begin{equation}
   \alpha w_7^{(0,1)} = -0.49 \, \text{GeV}^{-2}~,
\end{equation}
and thus about 25\% of the axial radius contribution to the pseudoscalar form factor. 
Including $\mathcal O(\alpha)$ corrections and evaluating the LO pseudoscalar form factor at $q^2 = -m_\mu^2$, we find that $w_7$ is given by
\begin{equation}
    w_7 = \left[40.02(60) - 0.49 \right]\, \text{GeV}^{-2}, 
\end{equation}
where the error is dominated by $g_{np \pi^\pm}$. 

\begin{figure}[b]
    \centerline{
    \includegraphics[width=0.3\columnwidth,clip=true]{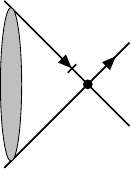}}
    \caption{Leading bound state muon-proton weak capture diagram in NREFT. Lines with (without) arrows are leptons (nucleons). A dash with an arrow indicates a non-relativistic lepton. The gray ellipse represents the Coulomb t-matrix.}
    \label{fig:boundstate}
\end{figure}

\section{Capture Rates}\label{sec:capturerates}
The general capture rate equation for non-relativistic nucleons and muon with relativistic neutrino is
\begin{equation}\label{eq:capturerate}
	\Gamma = \int \frac{d^3p_{\nu}}{(2\pi)^3} \frac{1}{2E_{\nu}} |\mathcal{M}(\mu p \rightarrow n + \nu_{\mu})|^2 (2\pi)\delta(\Delta E)~,
\end{equation}
where $\mathcal{M}(\mu p \rightarrow n + \nu_{\mu})$ is the invariant amplitude and $\Delta E = m_{\mu p} - E_{\nu} - \sqrt{m_n^2 + p_\nu^2} $, with $m_{\mu p}$ the mass of the bound muon-proton system. The set of Feynman diagrams for the bound state weak capture process are of the form seen in Fig.~\ref{fig:boundstate}.
We simplify Eq.~\eqref{eq:capturerate} by writing
\begin{align}
	\delta(\Delta E) &= \frac{\delta(E_\nu - E_{\nu 0})}{\left| 1 + \frac{E_{\nu 0}}{\sqrt{m_n^2 + E_{\nu 0}^2}} \right|} \nonumber\\
	&= \delta(E_\nu - E_{\nu 0}) \left( 1 - \frac{E_{\nu 0}}{m_{\mu p}} \right) \label{eq:deltafunc}~, 
\end{align}
where $E_{\nu 0}$ is the neutrino endpoint energy, 
\begin{equation}\label{eq:nuenergy}
E_{\nu 0} = \frac{m_{\mu p}^2 - m_n^2}{2m_{\mu p}}~.
\end{equation}
Expectedly, the capture rates are proportional to the muonic hydrogen wavefunction-at-origin squared. For some details of how that emerges, as well as some details in deriving the capture rate expressions, see Appendix~\ref{appendix:spinsums}. We include proton size effects to the wavefunction-at-origin so that~\cite{PhysRevLett.99.032003}
\begin{equation}
    |\tilde{\psi}_{\mu p}(0)|^2 \approx \frac{m_r^3 \alpha^3}{\pi} (1 - 0.005)~,
\end{equation}
where $m_r$ is the proton-muon reduced mass. Letting $C = G_F V_{ud} g_V$, the muonic hydrogen singlet and triplet channel capture rates are calculated as
\begin{widetext}
    \begin{align}
	\Gamma({}^1 S_0) &= |\tilde{\psi}_{\mu p}(0)|^2 \frac{C^2}{2\pi} E_{\nu 0}^2 \left( 1 - \frac{E_{\nu 0}}{m_{\mu p}} \right) \left| \left( w_1 + 3 w_6 \frac{g_A}{g_V} \right)  + \left( \frac12 w_2 \frac{g_A}{g_V} + \frac12 w_4 + w_5 \right) \frac{E_{\nu 0}}{m_N}  - \left( w_7 + w_3 \frac{E_{\nu 0}}{2m_N} \right) E_{\nu 0}^2 \right|^2~, \\
    \nonumber \\
    \Gamma({}^3 S_1) &= |\tilde{\psi}_{\mu p}(0)|^2 \frac{C^2}{3\pi}  E_{\nu 0}^2 \left( 1 - \frac{E_{\nu 0}}{m_{\mu p}} \right) \left\{ \frac12 \left| \left( w_1 - w_6 \frac{g_A}{g_V} \right) + \left( \frac12 w_2 \frac{g_A}{g_V} + \frac12 w_4 - w_5  \right) \frac{E_{\nu 0}}{m_N} - \left( w_7 + w_3 \frac{E_{\nu 0}}{2m_N} \right) E_{\nu 0}^2  \right|^2 \right. \nonumber \\
	&\hspace{2cm} + \left. \left| \left( w_1 - w_6 \frac{g_A}{g_V} \right) + \left( \frac12 w_4 - \frac12 w_2 \frac{g_A}{g_V}  \right)\frac{E_{\nu 0}}{m_N }  + \left( w_7 + w_3 \frac{E_{\nu 0}}{2m_N} \right) E_{\nu 0}^2   \right|^2 \right\} ~.
\end{align}
\end{widetext}
The above expressions yield singlet and triplet capture rates
\begin{align}
    \Gamma({}^1 S_0) &= \finalsinglet~\rm{s}^{-1} ~, \\
    \Gamma({}^3 S_1) &= \finaltriplet~\rm{s}^{-1}~.
\end{align}
Breaking down the capture rates by source, we find that the most dominant uncertainty in each channel arises from the uncertainty on the axial radius:
\begin{align}
    \Gamma({}^1 S_0) &= 724.07 \pm (5.20)_{\langle r^2_A \rangle} \pm (1.15)_{g_A/g_V} \nonumber \\
    &\hspace{0.25cm} \pm \left( 1.12\right)_{\text{other}}~\rm{s}^{-1} ~, \\
    \Gamma({}^3 S_1) &= 11.55\pm (0.16)_{\langle r^2_A \rangle} \pm (0.02)_{g_A/g_V} \nonumber \\
    &\hspace{0.25cm} \pm \left( 0.09\right)_{\text{other}}~\rm{s}^{-1} ~.
\end{align}
We note that we have taken a conservative stand and increased the uncertainty on $\langle r^2_A \rangle$ to 25\% of its mean value, which is more in line with other quantifications of that uncertainty. Within both channels, the other uncertainties are highly dominated by the uncertainty in the leading pseudoscalar interaction held in $w_7$.

Parameterizing the capture rate in terms of electroweak form factors and including the contribution from radiative corrections through a correction factor taken from Ref.~\cite{PhysRevLett.99.032003}, Hill {\it et al.}~\cite{Hill:2017wgb} obtain 715.4 (6.9)~s${}^{-1}$ and 12.10 (52)~s${}^{-1}$ for the singlet and capture rate, respectively. This compares well with our results quoted above with uncertainties of comparable size. In addition, the contributions of radiative corrections to our singlet capture rate comes in at $2.46(4)\%$, in excellent agreement with the size of radiative corrections in Ref.~\cite{PhysRevLett.99.032003}, while the contributions from radiative corrections are $1.64(16)\%$ in the triplet channel. As well as agreement with existing theoretical results, our singlet rate capture rate is in good agreement with the current best experimental determination from the MuCap collaboration, $715.6 (7.4)~\text{s}^{-1}$~\cite{MuCap:2015boo}. Experimental measurements of the triplet capture rate are not presently available, owing largely to the triplet capture rate being shorter than the free muon decay rate by a factor of $10^4$.

An additional source of uncertainty quoted in Ref.~\cite{Hill:2017wgb} is a finite size effect expressed as the first moment of the charge density $\langle r \rangle$. In an NRQED approach this term arises from a resummation of operators that give momentum dependence proportional to $q^{2n}$ with $n>1$. Treating these operators perturbatively leads to power divergences that need to be absorbed through counterterms that receive their information from the same proton-neutron transition matrix element.

\section{Summary and Outlook}
\label{sec:summary}
In this work, we have calculated the muonic hydrogen singlet and triplet weak capture rates using a fully non-relativistic electroweak EFT. Our approach leads to a Lagrangian structured in terms of heavy-to-light currents, which are constructed up to $\mathcal{O}\left( \frac{1}{m_\mu}\right)$. By comparing non-relativistic triangle diagram amplitudes to their relativistic counterpart, radiative corrections to the non-relativistic electroweak coupling were determined; specifically calculating only those contributing at $\mathcal{O}(\alpha)$. Renormalization scale dependence in those corrections is absorbed through the matching scale dependence in $g_V$,
so that by construction our calculations are thereby renormalization scale independent. We have found good agreement with previous experimental and theoretical results but more importantly a comparable result for the relative size of the radiative correction factor.

Our results demonstrate the power of our heavy-light current approach and open the path towards the calculation of radiative corrections with energetic probes in non-relativistic few-body systems where calculations can be carried out with both pionless EFT and chiral EFT. However, this might also require a systematic inclusion of heavy-light two-body currents and will require additional matching calculations. A straightforward first calculation will involve muon-capture on the deuteron, a reaction simpler than the proton-proton fusion calculation considered in Ref.~\cite{Combes:ppfusion} as both particles in the outgoing channel are neutral. It will of course also be important to test this approach in heavier systems and applications such as superallowed $\beta$-decay where one however also needs to account for radiative corrections that include photons coupling to nucleons that are not involved in the weak process.

\begin{acknowledgments}
We would like to thank E. ~Epelbaum, M.~Hoferichter and U.~van~Kolck for extensive discussions on the definition of the pion-nucleon coupling at $\mathcal O(\alpha)$, and A.~Meyer for conversations on the extraction of the axial radius in Lattice QCD. 
This work was supported by the National Science Foundation under Grant No. PHY-2412612 (EC \& LP) and the US Department of
Energy under Contract Nos. DE-AC05-00OR22725 (LP).
EM is supported by the U.S. Department of Energy Office and by the Laboratory Directed Research and Development (LDRD) program of Los Alamos National Laboratory under project numbers 20250164ER and 20260246ER. Los Alamos National Laboratory is operated by Triad National Security, LLC, for the National Nuclear Security Administration of the U.S. Department of Energy (Contract No. 89233218CNA000001)
We acknowledge support from the DOE Topical Collaboration “Nuclear Theory for New Physics” award No. DE-SC0023663.

\end{acknowledgments}

\begin{appendix}
\section{Spin-Averaged Amplitudes}\label{appendix:spinsums}
In this appendix we will detail some intermediate steps in calculating the spin-averaged amplitudes for muon capture on the proton in the spin-singlet and spin-triplet configurations. 

To begin with, it is useful to note that the muon-proton initial bound state with total angular momentum $J_{\mu p}$ and spin angular momentum $S_{\mu p}$, in the center-of-mass frame, can be written as~\cite{ChenChen1972}
\begin{equation}
    \ket{\mu p(S_{\mu p})} = \sum_{s_\mu, s_p} C^{J_{\mu p}, S_{\mu p}}_{\frac12, s_\mu; \frac12, s_p} t_C(\bm{k}) \ket{\mu (s_\mu, \bm{k})} \ket{p (s_p, -\bm{k})}~,
\end{equation}
where $t_C(k)$ is the Coulomb t-matrix. That formulation produces the wavefunction-at-origin for all terms in the amplitude through the relation~\cite{ChenChen1972}
\begin{equation}\label{eq:wavefunction}
    \tilde{\psi}_{\mu p}(0) = \int \frac{d^3k}{(2\pi)^3}\frac{t_C(\bm{k})}{-B_{\mu p} - \frac{\bm{k}^2}{2m_R} + i\varepsilon}~.
\end{equation}
Thus, the overall factor $|\tilde{\psi}_{\mu p}(0)|^2$ in the capture rates is produced. 

With that formulation of the initial muon-proton bound state, we find it is convenient, before phase space integration, to set the neutrino three-momentum to $\bm{p}_\nu = |\bm{p}_\nu|\hat{z}$, consequently setting the neutrino spin to $-\frac12$. This allows us to make a simplification to the leptonic amplitudes. For the time-like term,
\begin{align}
    \bar{u}_\nu \Omega^0 \chi_\mu^{s_\mu} &= \sqrt{E_\nu} \begin{pmatrix} \chi_\nu^T & -\chi_\nu^T \frac{\bm{\sigma} \cdot \bm{p}_\nu}{E_\nu} \end{pmatrix} \begin{pmatrix} \sigma^0 \\ \sigma^0 \end{pmatrix} \chi_\mu^{s_\mu} \nonumber \\
	&= \sqrt{E_\nu} \left[ \chi_\nu^T \sigma^0 \chi_\mu^{s_\mu} - \chi_\nu^T \frac{\bm{\sigma} \cdot \bm{p}_\nu}{E_\nu} \sigma^0 \chi_\mu^{s_\mu} \right] \nonumber\\
    &\rightarrow 2  \sqrt{E_\nu} \chi_{\nu \downarrow}^T \sigma^0 \chi_\mu^{s_\mu}~,
\end{align}
and likewise for the space-like term,
\begin{equation}
    \bar{u}_\nu \Omega^k \chi_\mu^{s_\mu} \rightarrow 2  \sqrt{E_\nu} \chi_{\nu \downarrow}^T \sigma^k \chi_\mu^{s_\mu}~,
\end{equation}
where the neutrino spinor $\chi_{\nu \downarrow}$ is a two-component spinor with spin-down eigenvalue. Coached in that language, the individual amplitudes are handily written in terms of products of Dirac $\delta$-functions and are easily summed over. For example, the Fermi interaction term yields
\begin{align}
    \braket{n, \nu_\mu| \bar{\Psi}_\nu \Omega^0  \psi_\mu \mathcal V_0 |\mu p} &= \tilde{\psi}_{\mu p}(0) w_1(m_\mu)2  \sqrt{E_\nu} \nonumber \\
    & \hspace{0.5cm} \times \chi_{\nu \downarrow}^T \sigma^0 \chi_\mu^{s_\mu} (\chi^{s_n}_N)^T \chi^{s_p}_N \nonumber\\
    &= \tilde{\psi}_{\mu p}(0) w_1 2  \sqrt{E_\nu} \delta^{s_\mu, -\frac12}\delta^{s_n, s_p}~.
\end{align}
Above, $\chi_\mu$ and $\chi_N$ are two-component spinors for muons and nucleons, respectively. Thus the spin-averaged amplitudes are made straightforward to calculate.

\section{Pion-photon loop functions}\label{appendix:loopfunctions}

After tensor reduction, the diagrams in Fig. \ref{fig:pseudo}
depend on relativistic triangle and bubble scalar functions and on box, triangle and bubble integrals with one linear propagator (heavy baryon integrals). 
The relativistic bubble integrals are easy to compute, while the triangle scalar function can be extracted in closed form from \texttt{PackageX} \cite{Patel:2015tea,Patel:2016fam}. For $q^2 = - m_\mu^2$, we obtain
\begin{align}
    f_1(x_\mu) &=  - \frac{\pi^2}{8} + \frac{1}{2}  {\rm Li}_2\left(\frac{1-x_\mu}{1+x_\mu}\right) - \frac{1}{2} {\rm Li}_2\left(-\frac{1-x_\mu}{1+x_\mu}\right)~. 
\end{align}
The heavy baryon integrals can
be written as one-parameter integrals of the function
\begin{equation}
    F(z,z_\mu) =\frac{1}{\sqrt{z (1-z)}}\left(\frac{\pi}{2} + \arctan \left(\frac{z_\mu-z }{\sqrt{z (1-z)}}\right)  \right)~,
\end{equation}
with $z_\mu = x_\mu/(1+x_\mu)$.
We will need the following functions that can be evaluated numerically:
    \begin{align}
    h_1(z_\mu) &=  \int_0^{z_\mu} dz F(z, z_\mu) ~,\\
h_2(z_\mu) &= \int_0^{z_\mu} d z \left(1 - \frac{z}{z_\mu}\right) F(z,z_\mu)  ~,  \\
h_3(x_\mu) &= \int_0^{z_{\mu}} d z \frac{z_{\mu} - z}{1- z} F(z, z_\mu)~.
\end{align}

\end{appendix}

\pagebreak

\end{document}